\documentclass[aps,floatfix,twocolumn,notitlepage,tightenlines,amsmath,amssymb,superscriptaddress,nofootinbib,11pt,longbibliography,pra]{revtex4-1}

\usepackage[utf8]{inputenc}
\usepackage[T1]{fontenc}
\usepackage[mathscr]{euscript}
\usepackage{graphicx}
\usepackage[dvipsnames]{xcolor}
\usepackage{amsthm}
\usepackage{psfrag}
\usepackage{ifsym}
\usepackage{dsfont}
\usepackage{enumerate}
\usepackage{amsfonts}
\usepackage{multirow} 
\usepackage{amsmath}
\usepackage{hyperref}
\hypersetup{colorlinks=true, linkcolor=red, citecolor=blue}
\usepackage[sort&compress]{natbib}
\usepackage[separate-uncertainty=true]{siunitx}
\usepackage{comment}
\usepackage[normalem]{ulem}
\usepackage{braket}

\begin{document}

\title{Experimental demonstration of optimal unambiguous two-out-of-four quantum state elimination}
\author{Jonathan W. Webb, Ittoop V. Puthoor, Joseph Ho, Jonathan Crickmore, Emma Blakely, Alessandro Fedrizzi \& Erika Andersson}

\affiliation{Institute of Photonics and Quantum Sciences, School of Engineering and Physical Sciences, Heriot-Watt University, Edinburgh EH14 4AS, United Kingdom}

\begin{abstract}
    A core principle of quantum theory is that non-orthogonal quantum states cannot be perfectly distinguished with single-shot measurements.
    However, it is possible to exclude a subset of non-orthogonal states without error in certain circumstances.
    Here we implement a quantum state elimination measurement which unambiguously rules out two of four pure, non-orthogonal quantum states---ideally without error and with unit success probability.
    This is a generalised quantum measurement with six outcomes, where each outcome corresponds to excluding a pair of states.
    Our experimental realisation uses single photons, with information encoded in a four-dimensional state using optical path and polarisation degrees of freedom. The prepared state is incorrectly ruled out up to $3.3(2)\%$ of the time.
\end{abstract}
\maketitle

\emph{Introduction.-}
Quantum state elimination (QSE)~\cite{Bandyopadhyay2014,BarnettBook,Wallden_2014,Heinosaari1, PhysRevResearch.2.013326,Crickmore2020} aims to exclude one or more quantum states from a given set.
Just like for the closely related task of quantum state discrimination~\cite{1969Qdae,IVANOVIC1987257,PERES198819,DIEKS1988303,Peres_1998,Barnett:09,Bergou2004}, QSE tasks can be optimised with regard to a number of competing payoffs: for example, minimum-error measurements always return an outcome but sometimes eliminate the wrong state; unambiguous measurements on the other hand never exclude the wrong state, but occasionally return an inconclusive result~\cite{Bandyopadhyay2014,Wallden_2014, Crickmore2020}.
Interestingly, for specific sets of states, QSE can satisfy both of these requirements
such that it is possible to always return an outcome without excluding the wrong state---this is the regime that our work focuses on.

A number of quantum information processing protocols~\cite{Perry2015, Heinosaari1, Heinosaari2, PhysRevResearch.2.013326} and communication schemes~\cite{Phoenix2000, Flatt2018} have been developed based on QSE.
Thus far QSE-based schemes have been realised using standard projective measurements.
For example, one-out-of-two quantum oblivious transfer---where a sender has two bits and a receiver obtains either the first or second bit in a way the sender does not know which bit was received---has been demonstrated in Ref.~\cite{Dusek_Wallden_Andersson_2021}.
Additionally, quantum digital signatures provide an information-theoretic secure method of verifying message authenticity amongst users, where projective elimination measurements remove the need for quantum memories~\cite{Collins2014}.
Investigations into the fundamental properties of the wavefunction also makes use of projective exclusion measurements~\cite{Pusey_2012}.

Eliminating non-orthogonal sets of states is in general more involved and often requires a generalised measurement.
An early related result was derived by Caves \emph{et al.}~\cite{Caves2002}, who found the sufficient conditions for when it is possible to unambiguously exclude one of three possible non-orthogonal quantum states with zero failure probability, as well as generalised exclusion of ``trine'' states---three states that are non-orthogonal and symmetric~\cite{BarnettBook, Phoenix2000, Flatt2018}.
In Ref.~\cite{Phoenix2000}, elimination measures on trine states are used such that the receiver can gain information through a broadcast post-measure of what state was not sent.

In this work, we experimentally realise an optimal quantum state elimination measurement that unambiguously rules out two of four non-orthogonal quantum states, as derived in Crickmore \emph{et. al}~\cite{Crickmore2020}.
A six-outcome positive operator valued measurement (POVM) is required where each outcome corresponds to eliminating a unique pair of states.
The POVM is mapped onto an optical platform and optimised by minimising the number of optical elements.
Our states are single photons encoded in the path and polarisation degree of freedom.
We prepare each of the four possible states, perform the state elimination measurement, and compare the results with theoretical predictions.

\emph{Theoretical framework.}-
Let us consider two qubits, each prepared in one of two states, $\ket{\pm\theta} = \cos\theta|0\rangle \pm \sin\theta |1\rangle$~\cite{Crickmore2020}.
The four possible two-qubit states can be summarised as,
\begin{align}
\label{eq:fourstates}
\ket{+\theta, \pm\theta} = & \cos^2\theta |00\rangle \pm \sin^2\theta|11\rangle \nonumber \\
                             & \pm \cos\theta\sin\theta ( |01\rangle \pm |10\rangle),\nonumber \\
\ket{-\theta, \pm\theta} = & \cos^2\theta |00\rangle \mp \sin^2\theta|11\rangle \nonumber \\
                            & \pm \cos\theta\sin\theta (|01\rangle \mp |10\rangle). 
\end{align}
When $\cos{2\theta}\le\sqrt{2}-1$, it is possible to exclude two of the four states.
When $\theta = \cos^{-1} \left( \sqrt{2}-1 \right)/2$, the states are the least distinguishable, but still such that two of them can be perfectly ruled out~\cite{Crickmore2020}. This is the case we consider.
From this set of four states, there are six ways to choose a pair of states.
Using the shorthand notation ``$++$'' for the state $\ket{+\theta,+\theta}$, ``$+-$'' for $\ket{+\theta, -\theta}$ and so on, the pairs we can eliminate are
\begin{eqnarray}
\label{eq:pairs}
A=\{++, +-\}, B=\{++, -+\}, \nonumber \\ 
C=\{+-, --\}, D=\{-+, --\}, \nonumber \\
E=\{+-, -+\}, F=\{++, --\}.
\label{eqn:pairs}
\end{eqnarray}

The generalised quantum measurement that achieves this can be realised by extending the four-dimensional Hilbert space to six dimensions, then making a projective measurement in the six-dimensional extended space.
The projective measurement in turn can be realised by applying a particular unitary transform $U$, followed by a projection in the computational basis~\cite{PhysRevA.63.052301}.
For details of this construction, see Supplemental Material.
In short, following Reck \emph{et al.}~\cite{Reck1994}, we decompose the unitary transform as a set of $2\times 2$ beam splitter-like (BS) operations.
This decomposition is optimised by permuting the ordering of the basis states through all possibilities, obtaining the decomposition in each case, and choosing the decomposition with the minimum number of BS operations.
We denote these BS operations by matrices $T_{ij}$ with $i,j \in \{1,...,6\}$.
One generally has $U \cdot T_{M,M-1} \cdot T_{M,M-2} \cdots T_{2,1} = D$, where $M$ is the number of outcomes, here $6$, and $D$ is a diagonal matrix.

\begin{figure}[t!]
    \includegraphics[width=\columnwidth]{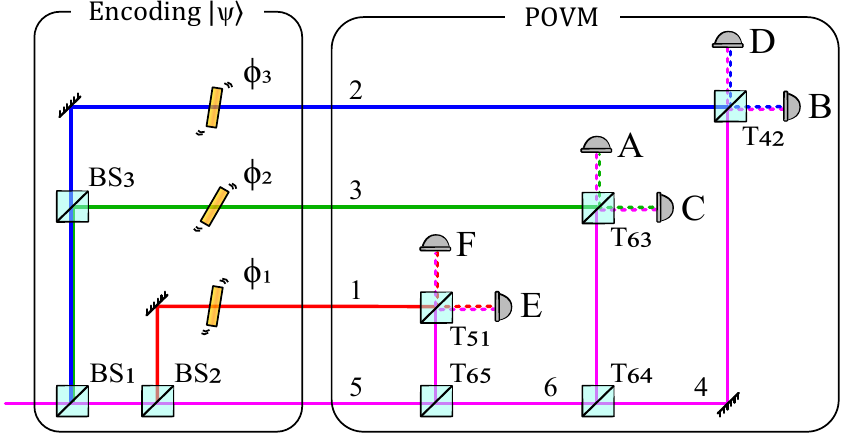}
    \caption{Two-out-of-four state elimination setup using linear optics. We encode the four quantum states $|\psi\rangle$ using $(BS_1, BS_2, BS_3)$ and $(\phi_1, \phi_2, \phi_3)$ which act on the four path modes corresponding to the basis states $|00\rangle, |01\rangle, |10\rangle, |11\rangle$. The state elimination POVM is realised by acting on the state encoded in the four path modes with the beam splitters $T_{ij}$ as described in the main text. Single-photon detectors monitor the six outcomes
    $A$, $B$, $C$, $D$, $E$ and $F$.}
    \label{fig:linoptsetup}
\end{figure}

We will present an optical implementation, as shown in Fig.~\ref{fig:linoptsetup}, but the method of constructing the physical realisation can be used also for other physical platforms. 
Here, the four basis states are realised using four separable path modes. 
The four quantum states defined in (\ref{eq:fourstates}) are encoded using BS$_{1,2,3}$ along with phase shifters ($\phi_{1,2,3}$).
After the optical network which realises the unitary transform $U$, a click in each of the detectors A-F corresponds to a pair of states being eliminated.
The states $|00\rangle, |11\rangle, |01\rangle$ and $|10\rangle$ are represented by modes $5, 1, 2$ and $3$ respectively.
The auxiliary states, $|\textrm{aux}_{1}\rangle$ and $|\textrm{aux}_{2}\rangle$, with vacuum states as input, are represented by modes $6$ and $4$.

\begin{figure*}[ht!]
    \centering
    \includegraphics[width=\textwidth]{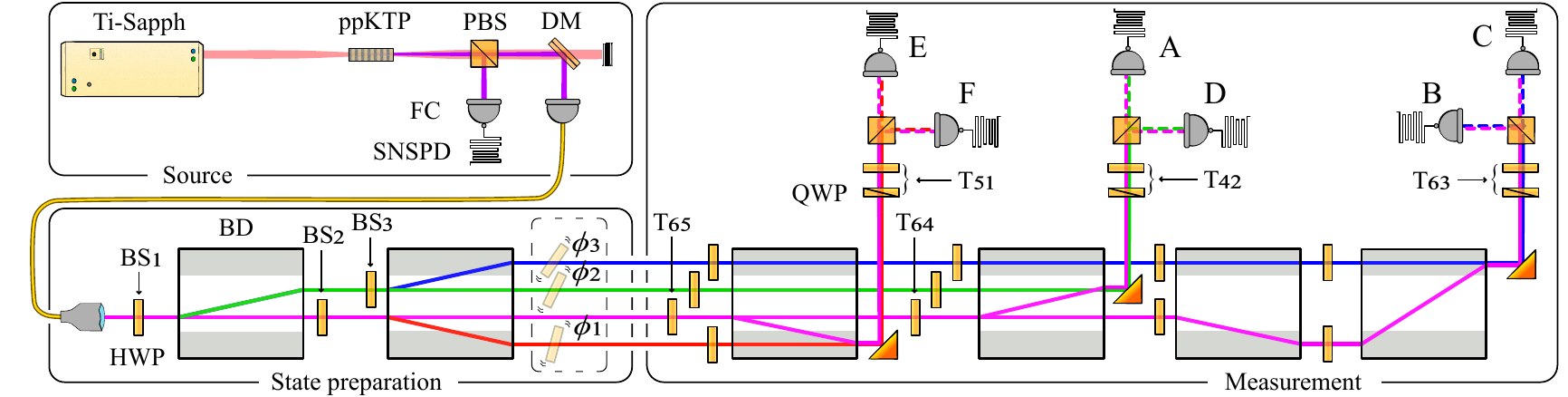}
    \caption{Experimental layout. A Ti-Sapph laser pumps a ppKTP crystal to produce a source of single photons.
    After a dichoric mirror (DM), single photons are fibre coupled (FC) to single mode fibre and sent to the state preparation stage.
    Four path modes are formed after two BDs along with HWPs to implement $\textrm{BS}_{1,2,3}$.
    The grey shaded BD areas denote paths that are vertically offset.
    The phases $\phi_{1,2,3}$, shown in the dashed box, are deferred to the measurement as described in the main text.
    The measurement uses HWPs and BDs to implement variable BS operations, $\{T_{65},T_{64}\}$, on the path modes. 
    Additional HWPs compensate path length differences.
    The remaining BS operations, $\{T_{51},T_{42},T_{63}\}$, are realised in polarisation through the use of a quarter-wave plate (QWP), HWP and PBS.
    Superconducting nanowire single-photon detectors (SNSPDs) detect the single photons in the six POVM output modes.
    }
    \label{fig:ExperimentSetup}
\end{figure*}

\emph{Experiment.}-
We encode the quantum states onto single photons produced by a heralded photon-pair source based on parametric down-conversion using a periodically-poled potassium titanyl phosphate (ppKTP) crystal~\cite{graffitti2018independent}.
Degenerate photon pairs are produced at 1550~nm through using a 775~nm pump from a 80~MHz pulsed Ti-Sapph laser.
The two down-converted photons are generated with orthogonal polarisation allowing them to be separated by a polarising beam splitter (PBS) and subsequently coupled into single mode fibre.
As shown in Fig.~\ref{fig:ExperimentSetup}, one of these photons is immediately detected as the herald photon while the other is sent to the state preparation stage.

We prepare the two-qubit states for the experiment by encoding the four-dimensional Hilbert space onto a single photon, effectively realising a ququart.
The ququart is encoded in the path-mode degree of freedom where coloured paths in Fig.~\ref{fig:linoptsetup} directly map to our experimental layout shown in Fig.~\ref{fig:ExperimentSetup}.
The four path modes are realised using two calcite beam displacers (BDs), which separates an input optical mode into two co-propagating paths depending on their polarisation.
BD interferometers generally have excellent passive phase stability and are widely used in multi-path and polarisation encoding in quantum optics setups~\cite{Xiao2018,Xiao2020,gao2020observation,PhysRevX.8.041007,Ringbauer_2015}.
By using half-wave plates (HWPs) on each path prior to the BD we set the relative splitting ratios of $BS_{1,2,3}$.
In the setup, the four paths co-propagate in a square lattice which is achieved by rotating the second BD by $90^\circ$ with respect to the propagation path, such that the spatial walk-off is in the vertical direction.
In the figure, the grey shading indicates the top two path modes while the unshaded regions represent the bottom two path modes.
Ideally, the phase shifts ($\phi_{1,2,3}$) are applied to the path modes directly as shown in Fig.~\ref{fig:ExperimentSetup} to prepare the different states in (\ref{eq:fourstates}), see Supplemental Material for details.
The unitary phase shifts $\phi$ commute with part of the unitary operation forming the POVM.
Hence, in the experiment these phases are implemented in the measurement stage, before the projective measurements and detection.

The POVM involves implementing $U$ on the path modes followed by a projective measurement in the extended basis spanned by the six outcomes.
As per the Reck~\emph{et al.} decomposition of $U$, we implement the network of beamsplitters, denoted by $T_{i,j}$, using both path and polarisation degrees of freedom.
We carry out part of the unitary transform on the path modes using BDs and HWPs, similar to the state preparation.
Specifically, the BDs spatially combine two path modes of orthogonal polarisation, with specific ratios set by the HWPs, into a single path mode.
The combined path modes retain orthogonal polarisation components which allows the final part of the unitary to be implemented in polarisation.
We use right-angle prism mirrors to direct the combined path modes towards polarisation measurement stages.
These consist of a QWP and HWP which implement the remaining $T_{i,j}$ operations in polarisation, as well as $\phi_{1,2,3}$, followed by a PBS for the projective measurement.

Finally, photons in each of the six path modes A-F, as defined in (\ref{eqn:pairs}), are detected with SNSPDs and a counting logic with a 1~ns coincidence window.
We conducted measurements with up to 30~s integration time for each experimentally prepared state.

\begin{figure}[ht]
    \includegraphics[width=\columnwidth]{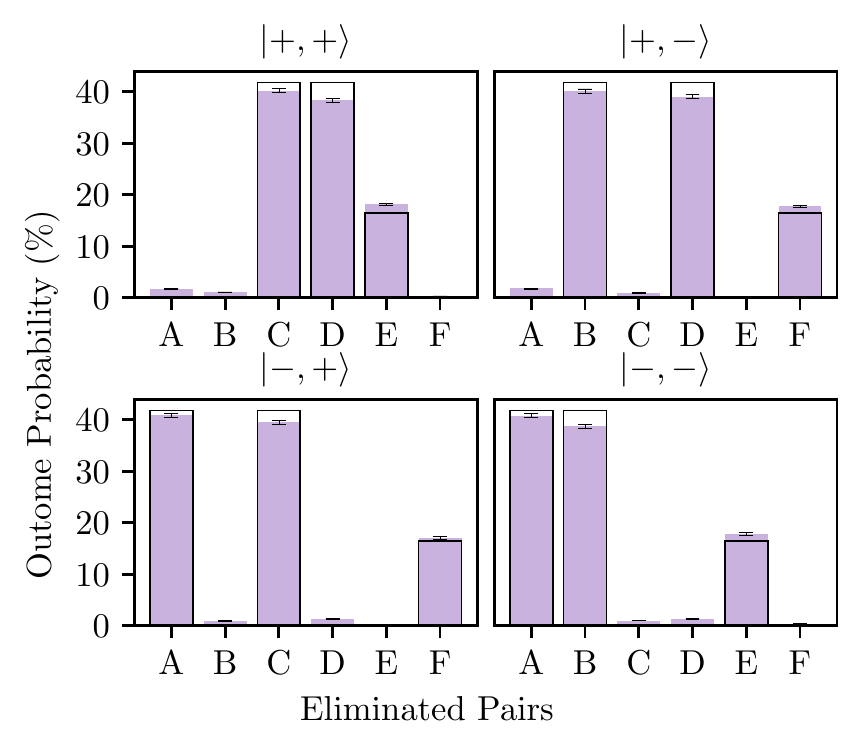}
    \caption{Experimental results (filled bars) for each of the states we prepared (\ref{eq:fourstates}). The theoretically predicted probability are indicated by a black outline.
    Each letter corresponds to a pair of states being excluded (\ref{eqn:pairs}).
    Error bars for the experimental data represent three standard deviations assuming Poissonian statistics.
    }
    \label{fig:ExperimentPlot}
\end{figure}

\emph{Results.}-
We prepared $\sim2.3$~x$10^{5}$ copies for each of the four states defined in (\ref{eq:fourstates}) and measured the output statistics of the POVM.
This is done by recording the number of detection events in each labelled path mode, A-F, which corresponds to eliminating a particular pair of states.
The experimental outcome probabilities are shown in Fig.~\ref{fig:ExperimentPlot} and compared with theoretical predictions.
For all states, mutually distinguishable sets of three outcomes are expected to have zero probability.
This is the signature that allows the unambiguous two-out-of-four state elimination.
Ideally the non-zero outcome probabilities for each state is $\sim\!42\%$ for outcomes A-D and $\sim\!16\%$ for E,F.
This distribution depends on $\theta$ which determines the set of states to be eliminated. If 
$\theta$ is any smaller than the value we have chosen, the task cannot be error-free.

The experimental data demonstrates good agreement with theoretical predictions, with a deviation of $1.3(2)\%$ on average across the entire data set.
Importantly, the task of state elimination was successful to within \{3.3(2)\%, 3.0(2)\%, 2.5(2)\%, 2.6(2)\%\} when preparing the states \{$\ket{++}$, $\ket{+-}$, $\ket{-+}$, $\ket{--}$\}, respectively.
Here, failure is considered as cases where outcomes that eliminate the state being prepared are detected.
The main contribution to imperfect elimination probabilities is non-unit visibility in the BD interferometers; on average we observe interference visibilities of 97.7\%.
This is somewhat lower than can be expected from BD interferometers, owing to the 2D layout of our optical path modes.

\emph{Discussion.}-
We have demonstrated the first two-out-of-four state elimination using a path mode encoding scheme for a specific set of states.
For both one- and two-out-of-four state elimination, 
measurements in an entangled basis are known to outperform local measurements (that is, separate measurements on each qubit)~\cite{Pusey_2012,Crickmore2020}.
Although our measurement would have required entangling operations if realized on two physically separate qubits, here it does not require entangling operations, because we have realised it using a ququart.
If one wants to maximise the average number of eliminated states from a set of qubits, local measurements are optimal~\cite{Crickmore2020}.
Unlike previous state elimination measurements, our demonstration realises a generalised unambiguous measurement. The total probability for an erroneous result, that is, to exclude a state that was prepared, happens up to $3.3(2)\%$ of the time.

Quantum state elimination is less explored than state discrimination, however it has already proven useful as a tool for a number of quantum information tasks~\cite{Collins2014,Phoenix2000,Dusek_Wallden_Andersson_2021}.
Conceivably, state elimination may enable a modified quantum key distribution (QKD) scheme where the secret key is created after the measurement is resolved -- similar to a prior QKD scheme that employs elimination of one among three states~\cite{Phoenix2000}.
That is, the sender prepares one of four non-orthogonal states, representing two bit values, then the recipient excludes two of these states, finally the two parties reconcile one secret bit: the first, second or XOR of the bits. This QKD scheme is conceptually different from traditional QKD therefore worthy of further study.
Theoretically, it's been shown that communication games based on excluding information through state elimination can be `infinitely' more efficient when using a quantum resource rather than a classical resource~\cite{Perry2015}.
For our specific scenario, weak-duality games can be demonstrated where two-out-of-four state elimination is required for a winning strategy~\cite{Hillery_2021}.
A future direction will be to expand the known strategies for eliminating optimal numbers of states in arbitrary state spaces. One approach is to use group theory to study state elimination measurements~\cite{hillerygroup}.

\vfill\eject

\textbf{Acknowledgements}

We thank Neil Ross for helping design and 3D print components. This work was supported by the EPSRC Quantum Communications Hub (grant No. EP/T001011/1.). JC acknowledges support from the EPSRC Condensed Matter Centre for Doctoral Training, grant No. EP/L015110/1.

\appendix
\begin{widetext}
\section{Realising the POVM}
Here we present the details of how we realised the POVM. The mathematical description of the optimal measurement is given in Crickmore \emph{et al.}~\cite{Crickmore2020}. When $\cos{2\theta}=\sqrt 2-1$, meaning that $2\theta \approx 65.5^\circ$, the measurement operators that eliminate each pair of states are given by $\Pi_i = |\psi_i\rangle \langle \psi_i|$, with $i \in \{A,B,C,D,E,F\}$.
Here, $|\psi_i\rangle$ are six unnormalised states with $\langle\psi_i|\psi_i\rangle<1$, which are not orthogonal to each other. Each state $|\psi_i\rangle$ is however orthogonal to the pair of states that are excluded when obtaining outcome $i$. We will realise the pair elimination measurement for this particular value of $\theta$. (When $\cos{2\theta}<\sqrt 2-1$, some of the measurement operators are mixed.)

Explicitly, we have
\begin{eqnarray}
\label{eq:orth1}
|\psi_{A}\rangle &= \frac{1}{\sqrt{2}}(\sqrt{\sqrt{2}-1} |00\rangle -|10\rangle), \nonumber \\ 
|\psi_{B}\rangle &= \frac{1}{\sqrt{2}}(\sqrt{\sqrt{2}-1} |00\rangle -|01\rangle), \nonumber \\ 
|\psi_{C}\rangle &= \frac{1}{\sqrt{2}}(\sqrt{\sqrt{2}-1} |00\rangle +|10\rangle), \nonumber \\
|\psi_{D}\rangle &= \frac{1}{\sqrt{2}}(\sqrt{\sqrt{2}-1} |00\rangle +|01\rangle), \nonumber \\ 
|\psi_{E}\rangle &= \frac{1}{\sqrt{2}}(\sqrt{2}-1 |00\rangle +|11\rangle), \nonumber \\ 
|\psi_{F}\rangle &= \frac{1}{\sqrt{2}}(\sqrt{2}-1 |00\rangle -|11\rangle).
\end{eqnarray}

\subsection{Completing the matrix}
In order to realise this generalised state elimination measurement, we need to translate the above mathematical description of the measurement into a physical setup. The first step is to employ the so-called Neumark extension. In short, this means that a generalised quantum measurement can be realised as a projective measurement in some higher-dimensional Hilbert space; we need as many dimensions as there are outcomes. We first write $|\psi_i\rangle = a_{r,1}|00\rangle + a_{r,2}|01\rangle + a_{r,3}|10\rangle + a_{r,4}|11\rangle$, where $r=1$ for $i=A$, $r=2$ for $i=B$, and so on, until $r=6$ for $i=F$.
We then form a $6\times 4$ matrix $V$ with elements $V_{r,j} = a^*_{r,j}$,
\begin{align}
\label{U4of6}
V= \frac{1}{\sqrt{2}}\begin{pmatrix}
\sqrt{\sqrt{2}-1} & 0 & -1 & 0 \\
\sqrt{\sqrt{2}-1} & -1 & 0 & 0 \\
\sqrt{\sqrt{2}-1} & 0 & 1 & 0 \\
\sqrt{\sqrt{2}-1} & 1 & 0 & 0\\
\sqrt{2}-1 & 0 & 0 & 1 \\
\sqrt{2}-1 & 0 & 0 & -1 
\end{pmatrix}.
\end{align}
Why we are doing this will become clear shortly. In the matrix $V$, the coefficients in each row belong to an unnormalised ``measurement state" $|\Psi_i\rangle$, and each column contains the coefficients for one particular basis state.

The completeness condition $\sum_i \Pi_i = 1$ ensures that the probabilities for all results sum to 1, no matter what state is being measured. This condition means that it holds that  $\sum_r a^{*}_{r,j}a_{r,j} = 1$ and $\sum_r a^{*}_{r,k}a_{r,l} = 0$, where $j,k,l \in \{1,2,3,4\}$ and $k \neq l$. This in turn means that the four columns in the matrix $V$ in equation \eqref{U4of6} are four six-dimensional orthonormal vectors. We can then complete the matrix into a $6\times 6$ unitary matrix by adding two more columns, corresponding to six-dimensional vectors which are orthonormal both to each other and to the four existing columns. That is, adding coefficients for two more auxiliary basis states $|aux_1\rangle, |aux_2\rangle$. Physically, the auxiliary basis states can be additional degrees of freedom for the existing quantum system(s), or can result from adding one or more auxiliary quantum systems; just a single auxiliary qubit would double the number of basis states. The mathematical description of how the measurement is realised is the same, but the physical realisation is different. 
In our realisation, the four original basis states and the two auxiliary states will all be represented by different spatial modes of a single photon.

If there are two or more columns to add, then this can be done in infinitely many ways. We can for example choose
\begin{eqnarray}
\label{Uopt}
U= 
\begin{pmatrix}
 &  & &  |& \frac{1}{2}  & \frac{1}{2}-\frac{1}{\sqrt{2}}\\
 &  & &  |& -\frac{1}{2} & \frac{1}{2}-\frac{1}{\sqrt{2}}\\
 &  \multirow{2}{*}{$V$} & & |& \frac{1}{2}& \frac{1}{2}-\frac{1}{\sqrt{2}}\\
 &  &  & |& -\frac{1}{2}& \frac{1}{2}-\frac{1}{\sqrt{2}}\\
 &  &  &  |& 0 & \sqrt{\sqrt{2}-1}\\
 &  &  &  |& 0 & \sqrt{\sqrt{2}-1}\\
\end{pmatrix}.& 
\end{eqnarray}
 No matter exactly how we choose to complete the unitary matrix $V$, to give a particular $U$, it then holds that 
\begin{equation}
    \langle i |U\rho U^\dagger |i\rangle = \langle \psi_i | \rho |\psi_i\rangle = {\rm Tr}(\Pi_i\rho)=p_i,
\end{equation}
where $|i\rangle$ is the basis vector corresponding to the $i^{\rm th}$ basis state, and $\rho$ is a state with support only in the four-dimensional ``original" Hilbert space.
This means that the measurement can be physically realised by performing $U$ on the original system, coupling it to the auxiliary degrees of freedom, followed by making a projective measurement in the basis given by the four original basis states plus the two auxiliary ones. 

\subsection{Decomposing the unitary operation into optical elements}
We search for a decomposition of $U$ containing the smallest possible number of $T$ matrices. By permuting input and output basis states before decomposing $U$ using the method in~\cite{Reck1994}, decompositions with different numbers of $T$ matrices can be obtained.  We look for a realisation with as few optical elements as possible. We denote the matrix obtained from $U$ by permuting input and output basis states, giving the smallest number of optical elements, by $U_{opt}$. The decomposition is then determined by $ D = U_{opt}  T_{65}  T_{64}  T_{63}  T_{51}  T_{42},$ where $D$ is a diagonal matrix and the remaining $T$ matrices are identity operations. We have
\begin{align}
U_{opt} =
\begin{pmatrix}
-\frac{1}{\sqrt{2}} & 0 & 0  & 0 & \sqrt{\sqrt{2} - 1} & 1 - \frac{1}{\sqrt{2}}\\
0 & -\frac{1}{\sqrt{2}} & 0  & -\frac{1}{2} & \frac{1}{2} - \frac{1}{\sqrt{2}} & \sqrt{\frac{1}{\sqrt{2}} - \frac{1}{2}}\\
0 & 0 & -\frac{1}{\sqrt{2}}  & \frac{1}{2} & \frac{1}{2} - \frac{1}{\sqrt{2}} & \sqrt{\frac{1}{\sqrt{2}} - \frac{1}{2}}\\
0 & \frac{1}{\sqrt{2}} & 0  & -\frac{1}{2} & \frac{1}{2} - \frac{1}{\sqrt{2}} & \sqrt{\frac{1}{\sqrt{2}} - \frac{1}{2}}\\
\frac{1}{\sqrt{2}} & 0 & 0  & 0 & \sqrt{\sqrt{2} - 1} & 1 - \frac{1}{\sqrt{2}}\\
0 & 0 & \frac{1}{\sqrt{2}}  & \frac{1}{2} & \frac{1}{2} - \frac{1}{\sqrt{2}} & \sqrt{\frac{1}{\sqrt{2}} - \frac{1}{2}}
\end{pmatrix}. 
\end{align}
The $T$ matrices are given by
\begin{align*}
T_{65} =
\begin{pmatrix}
1 & 0 & 0  & 0 & 0 & 0\\
0 & 1 & 0  & 0 & 0 & 0\\
0 & 0 & 1  & 0 & 0 & 0\\
0 & 0 & 0  & 1 & 0 & 0\\
0 & 0 & 0  & 0 & 0.91 & -0.414\\
0 & 0 & 0 & 0 & 0.414 & 0.91
\end{pmatrix}, 
T_{63} =
\begin{pmatrix}
1 & 0 & 0  & 0 & 0 & 0\\
0 & 1 & 0  & 0 & 0 & 0\\
0 & 0 & \frac{1}{\sqrt{2}}  & 0 & 0 & \frac{1}{\sqrt{2}}\\
0 & 0 & 0  & 1 & 0 & 0\\
0 & 0 & 0  & 0 & 1& 0\\
0 & 0 & -\frac{1}{\sqrt{2}} & 0 & 0 & \frac{1}{\sqrt{2}}
\end{pmatrix}, 
\:
T_{42} =
\begin{pmatrix}
1 & 0 & 0  & 0 & 0 & 0\\
0 & \frac{1}{\sqrt{2}} & 0  & \frac{1}{\sqrt{2}} & 0 & 0\\
0 & 0 & 1  & 0 & 0 & 0\\
0 & -\frac{1}{\sqrt{2}} & 0  & \frac{1}{\sqrt{2}} & 0 & 0\\
0 & 0 & 0  & 0 & 1 & 0\\
0 & 0 & 0 & 0 & 0 & 1
\end{pmatrix}, 
\end{align*}
\begin{align}
T_{64} =
\begin{pmatrix}
1 & 0 & 0  & 0 & 0 & 0\\
0 & 1 & 0  & 0 & 0 & 0\\
0 & 0 & 1  & 0 & 0 & 0\\
0 & 0 & 0  & \frac{1}{\sqrt{2}} & 0 & \frac{1}{\sqrt{2}}\\
0 & 0 & 0  & 0 & 1 & 0\\
0 & 0 & 0 & -\frac{1}{\sqrt{2}} & 0 & \frac{1}{\sqrt{2}}
\end{pmatrix}, 
\: 
T_{51} =
\begin{pmatrix}
-\frac{1}{\sqrt{2}} & 0 & 0  & 0 & \frac{1}{\sqrt{2}} & 0\\
0 & 1 & 0  & 0 & 0 & 0\\
0 & 0 & 1  & 0 & 0 & 0\\
0 & 0 & 0  & 1 & 0 & 0\\
-\frac{1}{\sqrt{2}} & 0 & 0  & 0 & -\frac{1}{\sqrt{2}}& 0\\
0 & 0 & 0 & 0 & 0 & 1
\end{pmatrix}. 
\end{align}

\section{Experimental details}
Details of how the experiment was performed are presented, starting from how we prepared the initial state.

\subsection{State preparation}
The states to be prepared were given in (\ref{eq:fourstates}), that is, they are
\begin{eqnarray}
\ket{+\theta, \pm\theta} = \cos^2\theta |00\rangle \pm \sin^2\theta|11\rangle \pm \cos\theta\sin\theta ( |01\rangle \pm |10\rangle),\nonumber\\
\ket{-\theta, \pm\theta} = \cos^2\theta |00\rangle \mp \sin^2\theta|11\rangle \pm \cos\theta\sin\theta (|01\rangle \mp |10\rangle).
\end{eqnarray}
To see how these states are prepared in our implementation, consider Fig.~\ref{fig:linoptsetup}, where each basis state $|00\rangle ,|01\rangle, |10\rangle, |11\rangle$ represents one of the coloured four path modes. The blue line represents the state $|01\rangle$, green $|10\rangle$, red $|11\rangle$ and purple $|00\rangle$.  Phase shifts of $\pm 1$ only need to be applied the basis states $|11\rangle$, $|10\rangle$ and $|01\rangle$. The phase shifts are achieved by adding a phase change to the appropriate path in an interferometer in our setup. As the basis state $|00\rangle$ requires no sign change, no phase shift is needed on the purple mode.

Preparing one of the four states in (\ref{eq:fourstates}) is achieved by rotating a HWP, corresponding to a $PS$, by $\pi$, or not rotating it. The notation $\phi_i, \ i=\{1,2,3\}$ is the same as in Fig.~\ref{fig:linoptsetup}. Each state is prepared through applying a rotation in a certain combination outlined in Table~\ref{table:desiredstate}. As an example, if preparing $\ket{+\theta , +\theta}$, the phase shifts are all $+1$, that is, no phase shift is required. Hence $\phi_i, \ i=\{1,2,3\}$ impart no rotation.

\begin{table}[h!]
\begin{flushleft}
\centering
\scalebox{0.9}{
\begin{tabular}{l l l l} 
 \hline\hline
 Desired state & $\phi_1$ & $\phi_2$ & $\phi_3$ \\ [0.5ex] 
 \hline
 $|+\theta,+\theta\rangle$ & 0 & 0 & 0\\ 
 $|+\theta,-\theta\rangle$ & $\pi$ & 0 & $\pi$\\
 $|-\theta,+\theta\rangle$ & $\pi$ & $\pi$ & 0\\
 $|-\theta,-\theta\rangle$ & 0 & $\pi$ & $\pi$\\[1ex] 
 \hline
 \hline
\end{tabular}
}
\caption{Preparation of a state through applying a phase to an arm of each MZI.
}
\label{table:desiredstate}
\end{flushleft}
\end{table}

Experimentally, we implement this operation by pushing the phase shift from a separable operation as seen in Fig.~\ref{fig:linoptsetup} in main text to one combined with the $T_{51}, T_{42}$ and $T_{63}$ operation. This is done out of experimental convenience and is true to the Fig.~\ref{fig:linoptsetup} in main text. To see this, one could realise the same experimental set up but separate the $T$ operations from the $PS$ operations by introducing a QWP-HWP-QWP configuration, giving access to the entire Poincare sphere, prior to the HWP that makes up the projection along with the PBS, which would allow for explicit realisation of the $PS$ operation. The two path modes' polarisation are still individually accessible as they have yet to combine on the final HWP. Hence, the operation of the QWP and HWP prior to the PBS is sufficient to perform the $T$ operations along with the $PS$ operations.

In order for the states to be prepared correctly, and the probability amplitudes of a photon detection at each detector be balanced, BS operations are performed at the encoding stage. 
To be explicit consider the following example with reference to Fig.~\ref{fig:linoptsetup} in main text. The MZI made up of the red and purple coloured path modes need to be balanced in terms of probability. That being; the red path being made a transmission through $BS_1$ and reflection through $BS_2$, and the purple path being composed of transmission through $BS_1$ and $BS_2$, and reflection of $T_{65}$. The $BS$ and $T_{65}$ operations should have beam splitter ratios that direct some amount of probability to not only balance the arms, but also give up enough probability amplitude to the other two interferometers.
Our method of determining how to balance each interferometer, that being how to determine the transmission and reflection coefficients of each beam splitting operation, is found via mode expansion. The $T$ matrix operations were presented prior along with the $BS$ operations $BS_i$ where $i=\{1,2,3\}$ below allow us to model a condition where each interferometer is balanced.
\begin{align*}
BS_{1} =
\begin{pmatrix}
\sqrt{\cos ^4\theta + \sin ^4\theta} & -\sqrt{2}\cos\theta\sin\theta \\
\sqrt{2}\cos\theta\sin\theta & \sqrt{\cos ^4\theta +\sin ^4 \theta}
\end{pmatrix}, 
BS_{2} =\frac{1}{\sqrt{\cos ^4\theta + \sin ^4\theta}}
\begin{pmatrix}
\cos ^2 \theta & -\sin ^2\theta \\
\sin ^2\theta & \cos ^2\theta
\end{pmatrix}, 
\end{align*}
\begin{align}
BS_{3} =\frac{1}{\sqrt{2}}
\begin{pmatrix}
1 & -1\\
1 & 1
\end{pmatrix}. 
\end{align}

The theoretical values for the transmission and reflection coefficients are given in Table~\ref{table:BSapproxvalues}. This table gives the coefficients for the $T$ matrix operations that take from the purple path mode in Fig.~\ref{fig:linoptsetup} in main text. The transmission and reflection values can be adjusted depending on setup specific losses. 
These values are a starting point which do not take into account experimental losses, but to be further optimised dependant on the experimental set up.

\begin{table}[h!]
\begin{flushleft}
\centering
\scalebox{0.9}{
\begin{tabular}{l l l} 
 \hline\hline
 BS operation & Transmission (\%) & Reflection (\%)\\ [0.5ex] 
 \hline
 $BS_1$ & $58.58$ & $42.42$ \\ 
 $BS_2$ & $85.36$ & $14.64$ \\
 $BS_3$ & $50.00$ & $50.00$ \\
 $T_{65}$ & $82.81$ & $16.81$ \\
 $T_{64}$ & $50.00$ & $50.00$ \\[1ex] 
 \hline
 \hline
\end{tabular}
}
\caption{The theoretical ratios between transmission and reflection each variable BS operates on their respective path-modes.}
\label{table:BSapproxvalues}
\end{flushleft}
\end{table}

Using this table, we can examine the percentage of the total probability amplitude available dedicated to each MZI. To be explicit, by following the path using Fig.~\ref{fig:linoptsetup} in main text as a map, the total probability amplitude the arms of each MZI has is found by multiplying the probabilities in Table.~\ref{table:BSapproxvalues}. For example, the first MZI whose output ports are measured by detectors E and F are made up of the red and purple paths. To re-iterate, the red path is made from a transmission of $BS_1$ which has a transmission probability of $58.58\%$ and a reflection of $BS_2$ which has a reflection probability of $14.64\%$. This makes the red path have a probability amplitude of $8.58\%$ of the total incident amplitude. Each path with their associated probability amplitudes of the total incident amplitude are is presented in Table.~\ref{table:Pathmodeprobability} below.

\begin{table}[h!]
\begin{flushleft}
\centering
\scalebox{0.9}{
\begin{tabular}{l l l} 
 \hline\hline
 Detector set & Path (colour) & Probability amplitude\\ [0.5ex] 
 \hline
 \multirow{2}{*}{E, F} & Red & 8.58\% \\
  & Purple & 8.41\% \\ 
 \multirow{2}{*}{A, C} & Green & 21.21\% \\
 & Purple & 20.7\% \\
 \multirow{2}{*}{D, B} & Blue & 21.21\% \\
 & Purple & 20.7\% \\
 \hline
 \multicolumn{2}{l}{Total} & 100.81\%\\[1ex]
 \hline
 \hline
\end{tabular}
}
\caption{Table of probability amplitude distribution along each arm. The detector set E and F have a total probability of $16.99\%$ of the total incident probability amplitude. Detector sets A, C and B, D take $41.91\%$ of the total incident probability amplitude. We see a total percentage higher than $100\%$. These values are derived from Table.~\ref{table:BSapproxvalues} which are susceptible to rounding errors. Experimentally, we are trying to balance each arm in accordance with these values and so this table is used more of as a guide rather than a definite limit.}
\label{table:Pathmodeprobability}
\end{flushleft}
\end{table}

It can be seen on Fig.~\ref{fig:linoptsetup} that the red path mode, making up the interferometer with outcomes E and F, is formed purely off the common purple path.
This is the physical cause of why the outcome probability amplitudes are unequal across the non-zero outcomes. 
Table.~\ref{table:Experimentaluncertainties} presents our outcomes when each state is prepared, specifically the data that is represented in Fig.~\ref{fig:ExperimentPlot} in the main text.

\begin{table}[h!]
\begin{flushleft}
\centering
\scalebox{0.9}{
\begin{tabular}{l@{\hskip 0.5cm} l l l l l l@{\hskip 0.5cm} l} 
 \hline\hline
 \multirow{2}{*}{State} & \multicolumn{6}{l}{Probability outcome of each detector (\%)} & \multirow{2}{*}{Total probability error (\%)}\\ [0.5ex]
 & A & B & C & D & E & F & \\[0.5ex]
 \hline
 $|+\theta,+\theta\rangle$ & 1.82(8) & 1.09(6) & 40.2(4) & 38.4(4) & 18.2(3) & 0.35(4) & 3.3(2) \\
 $|+\theta,-\theta\rangle$ & 0.95(6) & 40.2(4) & 0.95(6) & 39.1(4) & 0.24(3) & 17.7(3) & 3.0(2)\\
 $|-\theta,+\theta\rangle$ & 40.9(4) & 1.00(6) & 39.5(4) & 1.29(7) & 0.24(3) & 17.1(3) & 2.5(2)\\
 $|-\theta,-\theta\rangle$ & 40.8(4) & 38.7(4) & 1.01(6) & 1.27(7) & 17.8(3) & 0.36(4) & 2.6(2)\\[1ex]
 \hline
 \hline
\end{tabular}
}
\caption{For each prepared state, the probability amplitudes recovered from the single photon detections are presented along with their uncertainty. This experimental measurement uncertainty for detectors A-F is given by three standard deviations of the detector sets' probability amplitudes assuming Poissonian statistics. The total probability of getting a wrong result is defined as the sum over the probability amplitudes of the detectors that are not meant to click for a given state.}
\label{table:Experimentaluncertainties}
\end{flushleft}
\end{table}

\subsection{Additional experimental details}
All path modes co-propagate in parallel through the BDs, minimising any mismatch in optical path length and providing passive phase stability on all path-modes.
In order to operate on the correct path-mode in a setup whose compactness is defined by the size of the BD diameter, holed HWPs are used with custom 3D printed mounts.
These plastic 3D prints are subject to large thermal fluctuations and cause drift in the interferometer, reducing the visibility of each interferometer over time.
We regard this as the more dominant source of error, which could be mitigated by improving the mounting material from plastic to a higher thermal resistant material.
We remark that the BD path mode separation and recombination are determined by the orthogonal polarisation components.
Importantly, this means that the two optical modes, which are spatially overlapped after the BD, remain distinguishable in the polarisation degree of freedom. 

Each MZI has an associated loss, which can be defined by the contrast, that is dependent on the PBS at that closes the MZI. This is also due to an accumulative build up on non-perfect fringe visibilities of the BDs. Table.~\ref{table:Contrastvalues} below is the contrast of each port associated with a detector.
\begin{table}[h]
\begin{flushleft}
\centering
\scalebox{0.9}{
\begin{tabular}{l l} 
 \hline\hline
 Detector & Contrast (\%)\\ [0.5ex] 
 \hline
 A & $98.4\%$ \\ 
 B & $96.4\%$\\
 C & $98.2\%$\\ 
 D & $97.0\%$\\
 E & $97.2\%$\\
 F & $98.8\%$\\[1ex] 
 \hline
 \hline
\end{tabular}
}
\caption{The contrast values for each port leading to a detector. These were found by coupling a $1550$nm diode laser into each of the ports and by projecting H and V polarisations through the PBS using a HWP, we can obtain the maximum and minimum values for each transmission and reflection port. By using $(I_{max}-I_{min})/(I_{max}+I_{min})$ where $I$ is the intensity, we obtain the contrast.}
\label{table:Contrastvalues}
\end{flushleft}
\end{table}

It should also be noted that manufacturing constraints of the BDs add to the technical limitations of our set up.
Ideally, BDs should be matched such that they are mined from the same vein.
Non-perfect alignment in the lattice structure leads to dephasing which will limit the contrast of the interferometer.

\end{widetext}

\end{document}